\documentclass[
pra,twocolumn,
superscriptaddress
]{revtex4-2}

\usepackage{amsmath,amssymb}
\usepackage{graphicx}
\usepackage{dcolumn}
\usepackage{bm}
\usepackage{quantikz}
\usepackage{comment}
\usepackage{hyperref}

\begin{document}

\title{Adaptive Relational Learning on Multi-instance Quantum Data\\ with Photonic Processors}

\author{Marcin Jastrzebski}
 \affiliation{School of Mathematical and Physical Sciences, University of Sheffield, Sheffield S10 2TN, United Kingdom}

\author{Shang Yu}
\affiliation{Blackett Laboratory, Department of Physics, Imperial College London, London SW7 2AZ, United Kingdom}

\author{Raj B. Patel}
\affiliation{Blackett Laboratory, Department of Physics, Imperial College London, London SW7 2AZ, United Kingdom}

\author{Oleksandr Kyriienko}
\affiliation{School of Mathematical and Physical Sciences, University of Sheffield, Sheffield S10 2TN, United Kingdom}

\date{\today}

\begin{abstract}
Loading multiple quantum states in parallel into a quantum machine learning (QML) model can unlock learning tasks where key information resides in the \emph{relations} between states rather than in individual states. We introduce an adaptive relational learning framework for such multi-instance quantum data that accesses pairwise and higher-order relations. Our model combines global measurements via \textsf{SWAP} or \textsf{CYCLE} tests for evaluating an $n$-state Bargmann invariant with shallow trainable transformations applied locally to each input state. We demonstrate the approach for continuous-variable (CV) photonic systems, which naturally provide access to quantum data and necessary computing operations. We solve tasks involving hidden relationship detection, geometric phase classification, and sensing in the presence of an unknown shared nuisance interaction. We benchmark the adaptive model against a non-adaptive ``measure-first'' approach based on continuous-variable classical shadows, and show that the cost of shadow estimation grows rapidly with $n$, while our model avoids this dependence. Already for $n=2$, we achieve perfect test accuracy $\mathcal{A}=1.0$ with $500$ inference shots, improving average test accuracy over the shadow-based method by $\Delta\mathcal{A}=0.15$ while using $100$ times fewer shots per data point. Our work opens routes to sensing and quantum-data applications where adaptive photonic QML can access relational features that are costly to recover with non-adaptive, measure-first models.
\end{abstract}

\maketitle

\section{\label{sec:intro} Introduction}

Many machine learning (ML) tasks can best be understood as processing \textit{ensembles} of objects, where relevant information can reside in the relations between individual instances~\cite{zaheer2017deep, lee2019set}. Canonical examples include contrastive learning~\cite{hadsell2006dimensionality, schroff2015facenet} and multiple instance learning~\cite{dietterich1997solving, ilse2018attention}. In contrastive learning, the data is intentionally arranged into ensembles by the user, which can be seen as a form of data augmentation. This allows one to prepare configurations that highlight differences and similarities between the original ``atomic'' data points. Multiple instance learning (MIL), on the other hand, concerns scenarios where only an ensemble (a ``bag'') of instances is given a label, without additional information about individual instances or their relations. This setting can also be referred to as working with weakly annotated data~\cite{carbonneau2018multiple}.

Quantum machine learning (QML) is a growing field of research in which quantum processing is incorporated into machine learning pipelines. A significant part of modern QML focuses on classical data encoded into parametrized quantum circuits~\cite{Mitarai_2018, Benedetti_2019, masotllima2025prospectsquantumadvantagemachine}, with applications including classification~\cite{farhi2018classification,Li2021rev}, generative modeling~\cite{Zoufal2019,Rudolph2022,kyriienko2022protocols,kasture2022protocols,recioarmengol2026train,kurkin2025universal,bako2025fermionic,tuysuz2026dqgm}, solving differential equations~\cite{kyriienko2021solving,paine2023quantum}, and anomaly detection~\cite{Herr2021,kyriienko2022unsupervised,wozniak2023quantum}. In this setting, the choice of data encoding is itself a central design ingredient, strongly influencing the expressive power and generalization properties of quantum models~\cite{SchuldSweke2021PRA,Caro2021encodingdependent,umeano2026apl}. At the same time, recent work has highlighted a tension between trainability and classical simulability, with several structured and trainable model families admitting efficient classical surrogates or dequantization~\cite{Cerezo_2025,gilfuster2025traindequantize,rudolph2023classicalsurrogates}, while suitably structured tasks can still separate adaptive quantum processing from non-adaptive classical post-processing~\cite{kyriienko2025advantage}. 

A qualitatively different regime emerges when the data to be processed are quantum from the outset, with a processor receiving quantum states directly as inputs. Learning directly from such \textit{quantum data} can provide strong advantages in sample complexity~\cite{Huang_2022,Huang_2022_experiments,khan2025quantum}, and arises naturally when quantum states are generated by physical experiments or as outputs of quantum algorithms. For example, solutions produced by quantum differential-equation solvers can themselves be analyzed as quantum data~\cite{williams2024addressing,williams2025vortex}. In this setting, access to several quantum states within a single learning instance opens the possibility of extracting information encoded in their relations, rather than in each state individually. Recent results further show that even minimal quantum memory enabling joint measurements on suitably related quantum states can yield exponential sample advantages for specific learning tasks~\cite{king2024minimalmemory}. More generally, collective access to multiple copies of the same quantum state can be strictly more powerful for certain tasks~\cite{Conlon_2023,Haah_2017}. This motivates learning architectures that can process multiple quantum inputs jointly.

In this work, we propose an adaptive relational-learning architecture for quantum data composed of multiple input states. Each of the $n$ quantum inputs undergoes a shallow trainable transformation locally, while the states are brought together only in the final global measurement, implemented through a \textsf{SWAP} or \textsf{CYCLE} test. The resulting model output is the $n$-state Bargmann invariant~\cite{Bargmann1964}, a quantity of well-established importance in quantum information~\cite{Wagner_2024, oszmaniec2024measuring}. For $n=2$, this accesses pairwise similarity, while for $n>2$ the Bargmann invariants provide learning features that depend jointly on multiple quantum states and can encode information beyond pairwise relations.

Related multi-copy QML hypothesis classes have been studied previously, including settings where collective access to several copies can provide advantages over single-copy models~\cite{larocca2022group_prx}. Here, we focus on a more structured and application-driven setting in which the label of a data point depends on relations among independently prepared quantum states, and trainable local processing is used to adapt how those relations are probed. Continuous-variable (CV) photonics provides a natural realization of this architecture: generation, manipulation, interference, and measurement of quantum optical states can be integrated within a common physical platform~\cite{lenzini2018integrated, Clark_2026}, and photonic implementations of the \textsf{CYCLE} test can directly estimate the required Bargmann invariants~\cite{novo2026nativelinearopticalprotocolefficient}. This creates a natural route to adaptive QML for quantum data native tasks in sensing, networking, and related photonic applications~\cite{khan2025quantum}. Recent experiments have further demonstrated that CV photonic processors can execute quantum learning protocols~\cite{nielsen2025variational}, including ones in which entanglement-assisted joint measurements yield substantial sample-complexity advantages over conventional measurement strategies~\cite{Liu_2025}.

A central question is whether this adaptivity is actually necessary. Alternative measure-first strategies perform a fixed quantum data acquisition stage and defer subsequent learning to classical post-processing~\cite{Huang_2022, Jerbi_2026, Cerezo_2025, lerch2024efficientquantumenhancedclassicalsimulation}. Such approaches can remove the need for repeated adaptive access to a quantum processor for important classes of learning models, while theoretical separations also show that there exist quantum data learning tasks for which fully adaptive protocols are more powerful~\cite{gyurik2023limitationsmeasurefirstprotocolsquantum}. Our setting provides a concrete way to probe this boundary: the adaptive model learns local transformations before evaluating a global relational observable \cite{Umeano2026PRA}, whereas a measure-first surrogate must recover the same multi-state information from a fixed set of classical measurement data. The latter requires orders of magnitude more shots, already for simplest tasks with $n=2$.

We demonstrate the approach on hidden-relation detection, geometric phase classification, and sensing (Sec.~\ref{sec:use_cases}). We then benchmark the adaptive model against a non-adaptive protocol based on continuous-variable classical shadows (Sec.~\ref{sec:shadows}), showing that the cost of estimating the relevant global observables grows with the number of input states. Finally, in Sec.~\ref{sec:implementation}, we present an implementation blueprint for the proposed approach on the time-bin photonic processor \textit{Clavina}~\cite{yu_extensible_2026}.


\section{\label{sec:mqil}Relational learning using the Bargmann invariant}

We consider inputs consisting of an ordered collection of $n$ independently prepared, unentangled quantum states. Our architecture applies a trainable transformation locally to each input state and subsequently extracts relational information through a collective measurement. In particular, we use the $n$-state Bargmann invariant \cite{Bargmann1964} as the relational observable. This construction is illustrated in Fig.~\ref{fig:general_circuit} and described below.

\subsection{Proposed learning model}
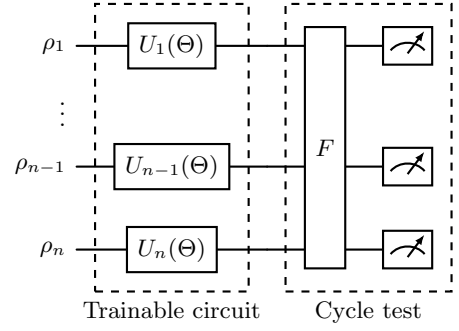
\begin{figure} \centering \begin{quantikz}[wire types={q,n,q,q}] \lstick{$\rho_1$} & \gate{U_1(\Theta)} \gategroup[4, steps=1, style={dashed}, label style={label position=below,anchor=north, yshift=-0.2cm}]{Trainable circuit}& &\gate[4]{F}\gategroup[4, steps=2, style={dashed}, label style={label position=below,anchor=north, yshift=-0.2cm}]{Cycle test} & \meter{} \\ \lstick{\vdots}&{} & & {} & {} & & \\ \lstick{$\rho_{n-1}$}& \gate{U_{n-1}(\Theta)} & & & \meter{} \\ \lstick{$\rho_n$} & \gate{U_{n}(\Theta)} & & & \meter{} \end{quantikz}
\caption{The proposed relational learning model. Each input state undergoes an independent trainable transformation, followed by a destructive \textsf{CYCLE} test. In photonic systems, the \textsf{CYCLE} test can be implemented using a Fourier interferometer, indicated here by the gate $F$. The measurement yields an estimate of the Bargmann invariant of the transformed input states.}
\label{fig:general_circuit}
\end{figure}

The model, pictured in Fig.~\ref{fig:general_circuit}, is defined by
\begin{equation}
\label{eq:model_definition}
f_{\Theta}(\rho_1,\dots,\rho_n)
=
\textrm{Tr}\left[
\textsf{CYCLE}
\bigotimes_{i=1}^{n}\rho_i(\Theta_i)
\right],
\end{equation}
where
\begin{equation}
\rho_i(\Theta_i)
=
U_i(\Theta_i)\rho_i U_i^\dagger(\Theta_i),
\end{equation}
and $\Theta$ denotes the collection of all local parameter sets $\Theta_i$. Importantly, the trainable transformations act independently on the $n$ input states and never entangle them. Relational information is instead accessed through the final collective \textsf{CYCLE} measurement.

For pure states $\rho_i=\ket{\psi_i}\bra{\psi_i}$, which we consider throughout this work, the $n$-state Bargmann invariant is defined as
\begin{equation}
\label{eq:bargmann_cycle_test}
\Delta_n
=
\textrm{Tr}[\rho_1\rho_2\dots\rho_n]
=
\langle \bigotimes_i^n \psi_i|
\textsf{CYCLE}
|\bigotimes_j^n \psi_j\rangle,
\end{equation}
and can therefore be interpreted as the expectation value of the \textsf{CYCLE} operator. For $n\geq3$, its value generally depends on the ordering of the input states, which allows the observable to retain information beyond unordered pairwise similarities.

For the special case $n=2$, the \textsf{CYCLE} operator reduces to \textsf{SWAP}, giving
\begin{equation}
\label{eq:swap_test}
\Delta_2
=
\langle \psi_1 \otimes\psi_2 |
\textsf{SWAP}
| \psi_1 \otimes\psi_2\rangle
=
|\langle\psi_1 | \psi_2 \rangle|^2,
\end{equation}
which is the squared overlap between the two states (or equivalently, their pure state fidelity). Efficient algorithms for estimating Eq.~\ref{eq:bargmann_cycle_test}, known as \textsf{CYCLE} tests, exist for both qubits~\cite{oszmaniec2024measuring} and photonic quantum systems~\cite{novo2026nativelinearopticalprotocolefficient}.

Already for $n=2$, independent trainable transformations allow the model to go beyond a fixed kernel-like state overlap. Consider an asymmetric ansatz in which a unitary $U_1$ is applied only to the first state. The model then evaluates
\begin{equation}
\label{eq:aswap}
\begin{aligned}
f(\ket{\psi_1},\ket{\psi_2})
&=
\langle \psi_1 \otimes \psi_2 |
(U_1^{\dagger} \otimes I)
\textsf{SWAP}
(U_1 \otimes I)
| \psi_1 \otimes \psi_2 \rangle \\
&=
\left|
\langle \psi_2 | U_1 | \psi_1 \rangle
\right|^2.
\end{aligned}
\end{equation}
Thus, the trainable transformation turns the fixed state overlap into a squared transition amplitude, allowing the model to act as a detector of unitary relationships between the two inputs.

A related interpretation arises when the two input states are identical, $\ket{\psi_1}=\ket{\psi_2}=\ket{\psi}$, but undergo different transformations $U_1(\Theta_1)$ and $U_2(\Theta_2)$. In this case,
\begin{equation}
\label{eq:sensing_is_loschmidt}
f(\ket{\psi}) = 
|\bra{\psi}U_1^\dagger U_2\ket{\psi}|^2 =
\mathcal{L}_{\ket{\psi}, U_1, U_2},
\end{equation}
where $\mathcal{L}_{\ket{\psi}, U_1, U_2}$ is the Loschmidt echo~\cite{Wisniacki_2012}. It quantifies the extent to which the action of one process can be reversed by another and provides a natural connection between the relational model and sensing tasks, where $\mathcal{L}$ is used, e.g., for signal amplification~\cite{Colombo_2022} and sensitivity-maximization~\cite{liu2025variational}.

For $n=2$, the Bargmann invariant is real and captures pairwise overlap information. From $n=3$ onward, it can in general be complex, with its phase providing access to higher-order relational information that is absent from pairwise fidelities~\cite{Wagner_2024}. A key physical quantity encoded in this phase is the geometric phase. For an ensemble of non-orthogonal states, it can be obtained from the argument of the Bargmann invariant~\cite{Rabei_1999}:
\begin{equation}
\varphi_g[\rho_1,\rho_2,\dots,\rho_n]
=
-\arg{\Delta_n(\psi_1,\psi_2,\dots,\psi_n)},
\end{equation}
for the corresponding path through ray space obtained by connecting successive states with so-called \textit{null-phase curves}.

In the remainder of this work, we develop and study this relational learning architecture using continuous-variable (CV) states. CV systems provide a natural setting for quantum sensing~\cite{Guo_2019}, networking~\cite{Usenko_2026}, and simulation~\cite{deng2016continuous, Abel_2024}, while also providing native optical implementations of the required collective measurements. Sample-based emulation results use the destructive \textsf{CYCLE} and \textsf{SWAP} test protocols for optics~\cite{novo2026nativelinearopticalprotocolefficient, Volkoff_2022}.


\section{\label{sec:use_cases}Examples and applications}

We consider supervised learning tasks on labeled quantum datasets
\begin{equation}
\mathcal{D}=\{(\ket{d_i},y_i)\}_{i=1}^{D},
\end{equation}
where each data point is a product of $n$ pure quantum states,
\begin{equation}
\ket{d_i}
=
\bigotimes_{j=1}^{n}\ket{\psi_j^{(i)}},
\end{equation}
and $y_i$ denotes the corresponding label. We assume access to repeated preparations of each data point, as required to estimate the relevant observables from a finite number of measurement shots. The associated shot requirements are discussed in detail in Sec.\ref{sec:shadows}.

For each task, we compare the relational model with a classifier using a local observable, namely the number operator $\hat{n}_1$ on the first mode, together with the same or a more expressive trainable gate set (further details are provided in Appendix~\ref{app:local_models}). This comparison is motivated by the well-known connection between global observables and barren plateaus in sufficiently expressive variational quantum circuits~\cite{Cerezo_2021}, which led to the prevalence of architectures with local observables. Here, instead, the global observable is combined with shallow, physics-informed ansatzes of limited expressivity, which do not experience reduced gradients.

Throughout this section, we use Gaussian states to facilitate simulation and visualization of the relational structure underlying each task. These examples are intended to illustrate the learning mechanisms rather than to establish an advantage over efficient Gaussian state-specific methods. The same relational constructions, however, apply more generally beyond the Gaussian setting.


\subsection{Hidden relationship detection}
\label{sec:hidden_relationship}
\begin{figure}[b!]
\centering
\includegraphics[width=\linewidth]{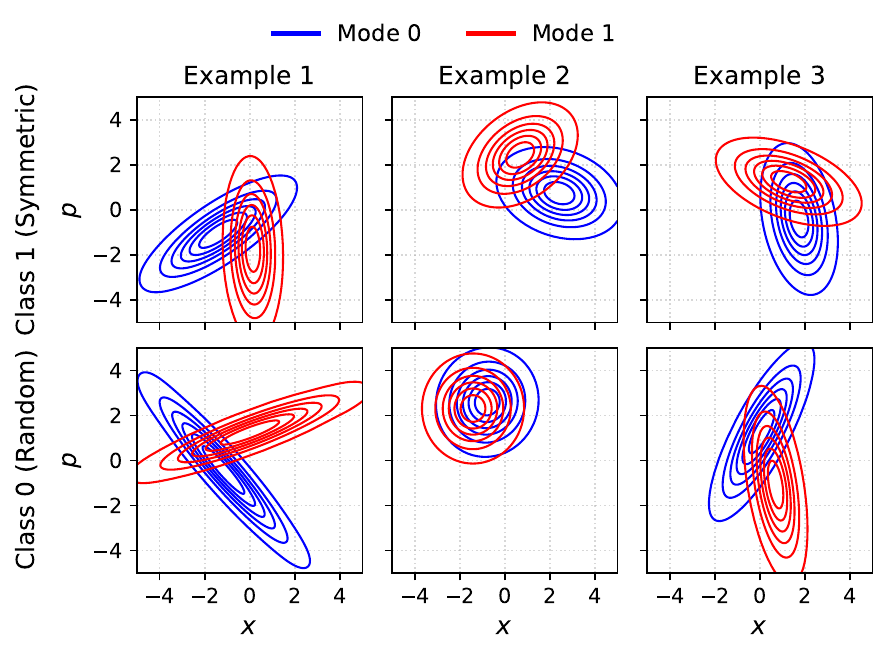}
\caption{Visualization of example pairs from the two classes of the hidden relationship problem. Class $1$ pairs (top) differ by a fixed relative rotation, whereas class $0$ pairs (bottom) differ by a random relative rotation.}
\label{fig:hidden_relationship_dataset}
\end{figure}

We first consider the pairwise case, $n=2$, where the Bargmann invariant reduces to the squared state overlap [Eq.~\ref{eq:swap_test}]. The task is to distinguish pairs of states related by a fixed but initially unknown rotation from pairs with a random relative rotation. Example data points are visualized in the Wigner plane in Fig.~\ref{fig:hidden_relationship_dataset}.
\begin{figure}[t!]
\centering
\includegraphics[width=\linewidth]{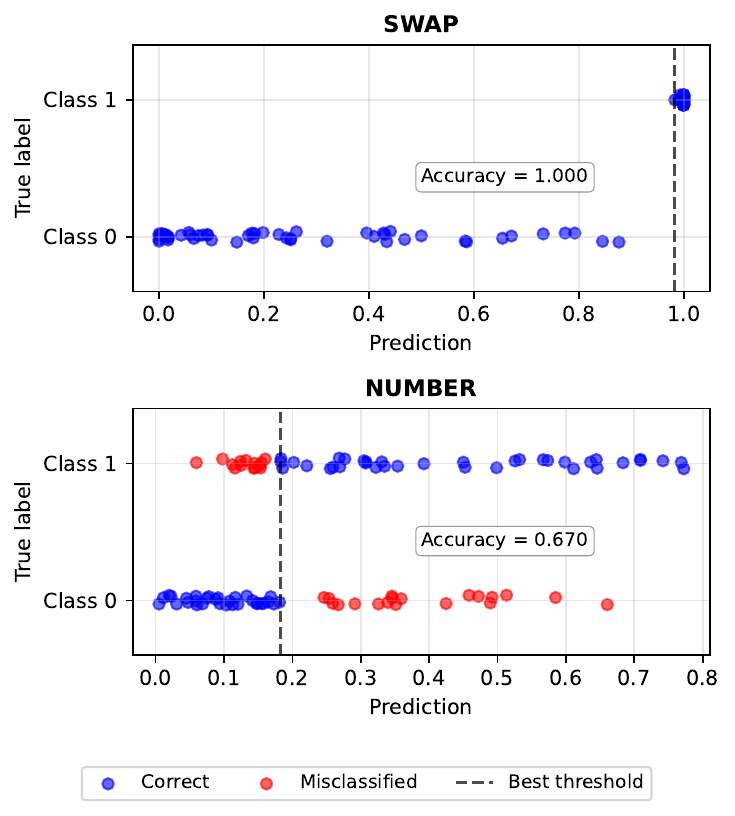}
\caption{Predictions on the test data for the hidden relationship problem using the \textsf{SWAP}-based relational model (top) and the local-observable model (bottom). Correctly classified points are shown in blue and misclassified points in red. The dashed black line denotes the classification threshold.}
\label{fig:rot_data_results}
\end{figure}

In this supervised binary classification task, each data point consists of two states,
\begin{equation}
\ket{\psi_i}
=
R(r_i)D(\alpha_i)S(s_i,\theta_i)\ket{0},
\qquad i\in\{1,2\},
\end{equation}
where $R$, $D$, and $S$ denote rotation, displacement, and squeezing, respectively, and $\ket{0}$ is the vacuum state. The displacement and squeezing parameters are sampled randomly and shared by the two states in each pair. The classes differ only in their relative rotation: for class $1$,
\begin{equation}
r_2-r_1=r_{\textrm{hidden}},
\end{equation}
whereas for class $0$ the relative rotation is sampled from
\begin{equation}
\mathcal{R}
=
[0,2\pi)
\setminus
[r_{\textrm{hidden}}-\epsilon,
r_{\textrm{hidden}}+\epsilon].
\end{equation}
A relational model containing a single trainable rotation $R(\theta)$ on one of the two inputs is sufficient to identify this hidden relationship. At the optimum, $\theta^*=r_{\textrm{hidden}}$, the class-$1$ pairs are aligned and therefore yield
\begin{equation}
f(\ket{\psi_1},\ket{\psi_2})=1,
\end{equation}
whereas class-$0$ pairs remain imperfectly aligned and satisfy
\begin{equation}
f(\ket{\psi_1},\ket{\psi_2})<1.
\end{equation}
The value of $\epsilon$ therefore determines the separation between the two classes and the precision with which $\theta^*$ must be learned.

Fig.~\ref{fig:rot_data_results} shows the results for $\epsilon=0.5$, using $20$ training instances and $50$ training epochs. Performance is evaluated on $100$ unseen data points and compared with the local-observable model described above. The local model reaches an accuracy of $\mathcal{A}=0.67$, whereas the \textsf{SWAP}-based relational model achieves perfect classification, $\mathcal{A}=1.00$.

We note that this Gaussian example is intended as a transparent demonstration of relational learning rather than as evidence of an advantage over efficient Gaussian state-specific methods. Collective comparison becomes particularly relevant when local characterization is costly, as can occur for general non-Gaussian states, or when the relation of interest is inaccessible from independent local measurements. The latter situation arises naturally when two sources do not share a common phase reference. Estimating and compensating such relative phase offsets is a central ingredient of twin-field QKD~\cite{Lucamarini_2018}. The \textsf{SWAP} test also has a direct connection to Hong-Ou-Mandel interference, a standard probe of photonic indistinguishability~\cite{Garcia_Escartin_2013}, and adaptive calibration of this quantity through measurement-driven optimization has been demonstrated previously~\cite{cortes2022sample}. Here, instead, the overlap serves as the output of a trained relational model.

\subsection{Chirality-based classification}
\label{sec:chirality}

We next consider the first genuinely higher-order case, $n=3$, where the Bargmann invariant can be complex and its phase carries information unavailable from pairwise overlaps alone. In particular, we study a classification task based on the connection between the three-state Bargmann invariant and the geometric phase (see Sec.~\ref{sec:mqil}). For coherent states, this phase admits a simple geometric interpretation in terms of the oriented area enclosed by the triangle formed by their phase-space centers. This example therefore highlights a key limitation of pairwise similarity measures: while pairwise overlaps encode distances between states, they do not capture the orientation of a trajectory in state space.
\begin{figure}[t!]
    \centering
    \includegraphics[width=\linewidth]{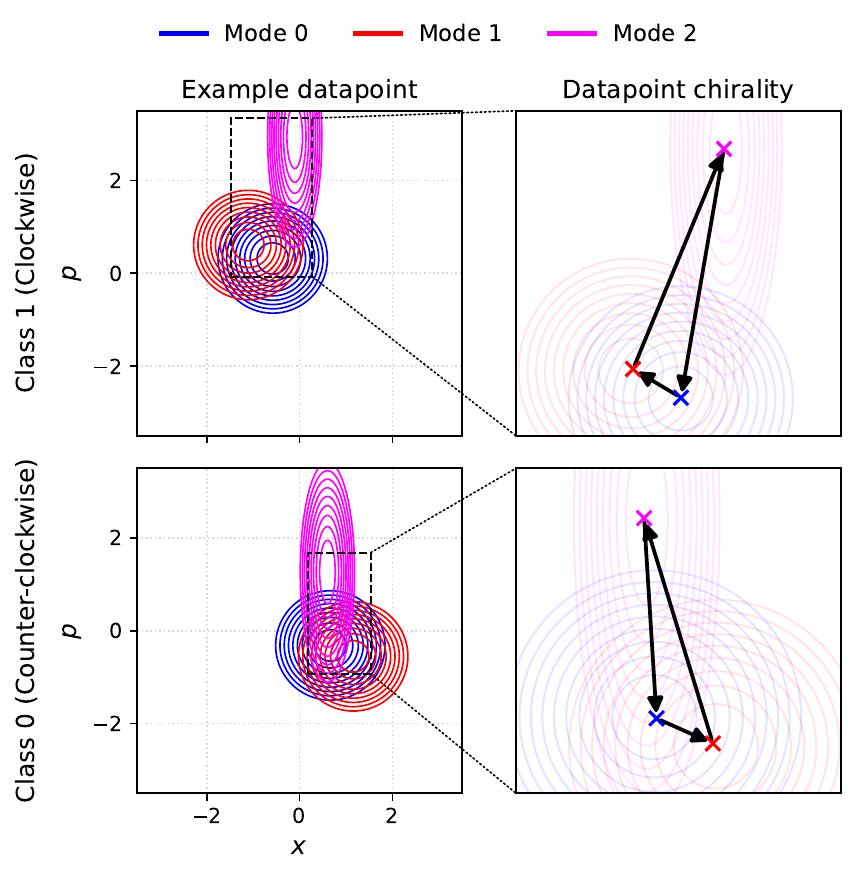}
    \caption{Visualization of example triplets from the two classes of the chirality-based classification task. Class $1$ (top) corresponds to clockwise orientation, whereas class $0$ (bottom) corresponds to anticlockwise orientation. Applying an unknown squeezing transformation to one state obscures the direct relation between this geometric picture and the Bargmann invariant.}
    \label{fig:chiral_dataset}
\end{figure}

The dataset is constructed from triplets of coherent states. The label assigned to each triplet is its chirality, namely the handedness of the closed loop passing through the three phase-space centers. The two classes thus correspond to opposite orientations of the triangle in the Wigner plane.

To make the task non-trivial, one state in each triplet is subjected to a fixed but unknown squeezing transformation before being presented to the model. This obscures the direct connection between the chirality of the underlying triplet and the phase of $\Delta_3$. As a result, the label can no longer be inferred reliably from the distorted states by simply inspecting the sign of $\arg \Delta_3$. Instead, a variational circuit is trained to compensate for the unknown distortion so that the phase of the resulting three-state Bargmann invariant again correlates with the label. In this way, the model learns to recover a geometric feature of the data that has been hidden by the squeezing operation.

\begin{figure}[b!]
    \centering
    \includegraphics[width=\linewidth]{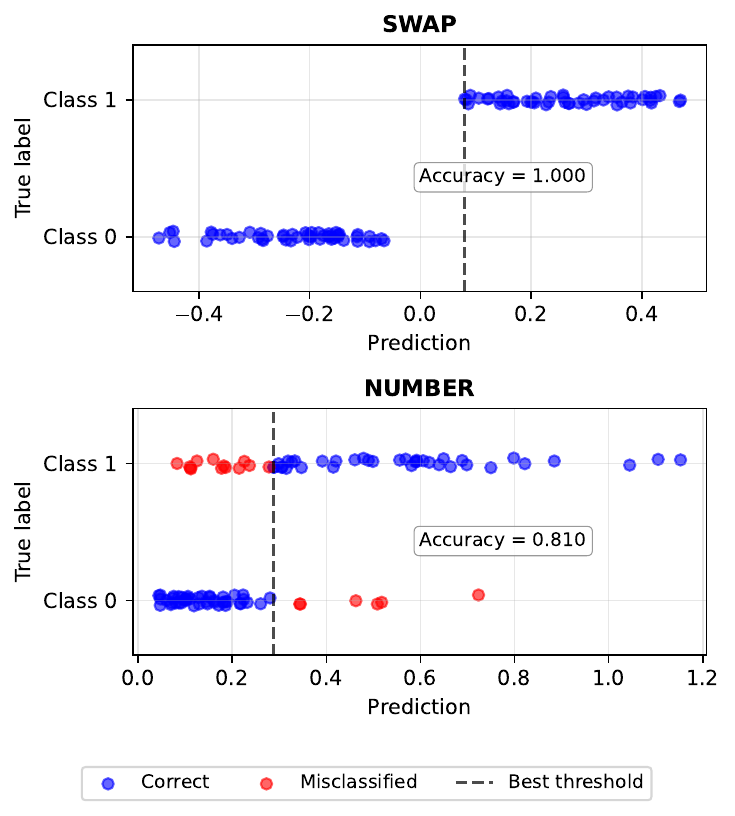}
    \caption{Predictions on the test data for the chirality-based classification task using the \textsf{CYCLE}-based relational model (top) and the local-observable model (bottom). Correctly classified points are shown in blue and misclassified points in red. The dashed black line denotes the classification threshold.}
    \label{fig:chiral_dataset_predictions}
\end{figure}

Examples of triplets from the two classes are shown in Fig.~\ref{fig:chiral_dataset}. The classification performance of the trained model is presented in Fig.~\ref{fig:chiral_dataset_predictions}. The model is trained on $20$ data points for $20$ epochs and evaluated on a held-out test set of $100$ points. The local-observable model reaches an accuracy of $\mathcal{A}=0.81$, whereas the \textsf{CYCLE}-based relational model again achieves perfect classification, $\mathcal{A}=1.00$. Again, this Gaussian example is intended primarily as a transparent illustration of how higher-order relational information, encoded in the phase of the Bargmann invariant, can be exploited for learning.


\subsection{Adaptive quantum sensing with biased probes}
\label{sec:sensing_task}

As a final example, we consider a quantum sensing task in which the probe is affected by an unknown quasistatic nuisance interaction after acquiring the signal of interest. We show that a two-copy relational measurement can cancel this common bias, allowing the model to classify the signal without estimating the nuisance interaction itself.

We consider a vacuum probe, $\ket{\psi_0}=\ket{0}$, and squeezing signals $G_i=S(z_i)$, where
$S(z)=\exp[(z^*a^2-z{a^\dagger}^2)/2]$, $z\in\mathbb{C}$, and $a$ ($a^\dagger$) denotes the bosonic annihilation (creation) operator. In the absence of additional noise, squeezing along orthogonal directions in phase space can be distinguished directly using homodyne measurements. We consider a binary classification problem in which the sensing interaction produces
\begin{equation}
\label{eq:sensing_data}
\ket{\psi_i}=G_i\ket{\psi_0},
\end{equation}
with labels $y_i\in\{0,1\}$. The two classes correspond to orthogonal squeezing directions: $z_i$ is imaginary for $y_i=0$ and real for $y_i=1$.

During the acquisition of each data point, the probe subsequently undergoes an unknown phase rotation $R(\phi_i)$. The angle $\phi_i$ may vary between data points but is assumed to remain fixed over the repeated preparations used to estimate a given data point. When this rotation is unknown and comparable in magnitude to the signal, it obscures the phase-space orientation that distinguishes the two signal classes.
\begin{figure}[h!]
\centering
\includegraphics[width=\linewidth]{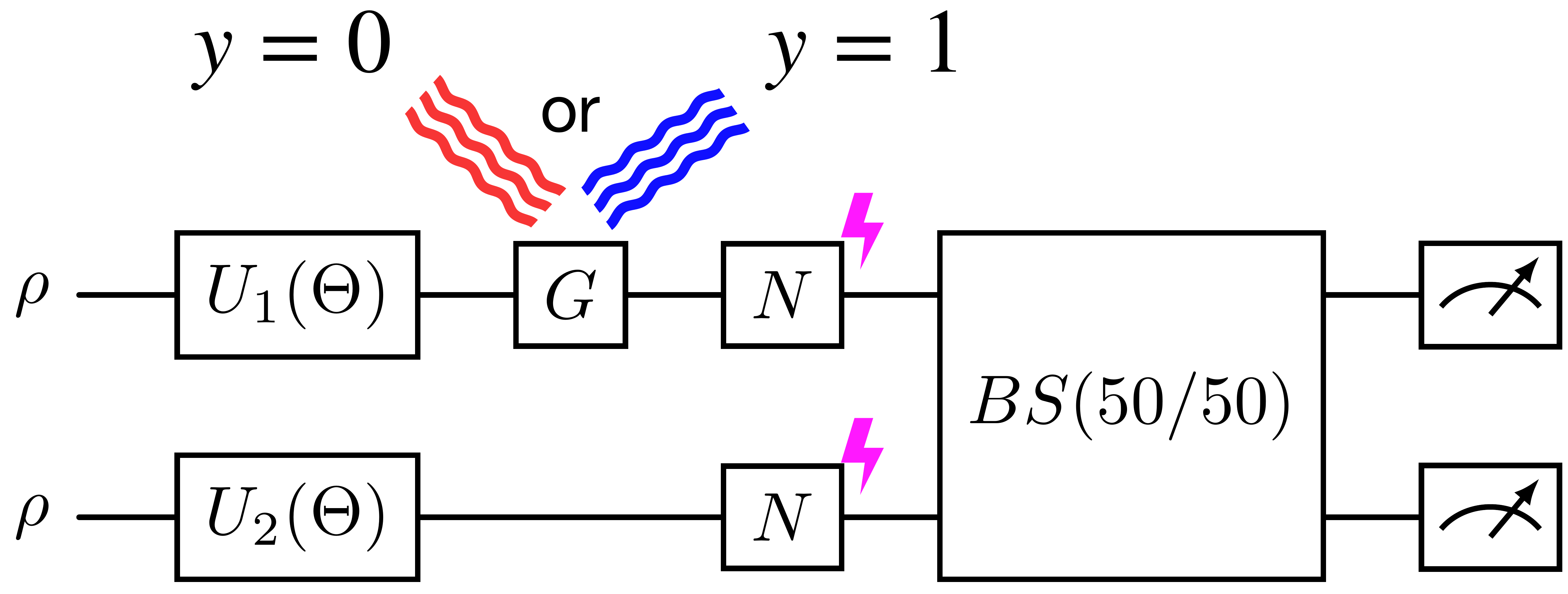}
\caption{Two-copy relational model for the quantum sensing task. One copy interacts with the signal $G_i$, while both copies undergo the same unknown nuisance interaction before being compared using a \textsf{SWAP} test. The model is trained to distinguish the two signal classes $y_i\in{0,1}$ while remaining insensitive to the shared nuisance interaction.}
\label{fig:sensing_task}
\end{figure}

To remove this nuisance dependence, we use the two-copy protocol shown in Fig.~\ref{fig:sensing_task}. Two copies of the probe are prepared and experience the same nuisance interaction, while only one copy interacts with the signal $G_i$. Independent trainable transformations $U_1$ and $U_2$ are initially applied to the two copies, which are compared using a \textsf{SWAP} test. The resulting model output is
\begin{equation}
\label{eq:sensing_nuisance_cancellation}
\begin{aligned}
f(\ket{\psi_i})
&= \left|\bra{\psi_0}U_1^\dagger G_i^\dagger
R^\dagger(\phi_i)R(\phi_i)U_2\ket{\psi_0}\right|^2 \\
&= \left|\bra{\psi_0}U_1^\dagger G_i^\dagger
U_2\ket{\psi_0}\right|^2 = \mathcal{L}_{\ket{\psi_0},\,G_iU_1,\,U_2}.
\end{aligned}
\end{equation}
Here, $\mathcal{L}$ denotes the Loschmidt echo introduced in Sec.~\ref{sec:mqil}. Because the same rotation acts on both copies, it cancels exactly from their overlap. More generally, the same cancellation occurs for any unknown unitary applied identically to both copies immediately before the relational measurement.

The learning task is therefore to choose the trainable transformations such that the nuisance-independent overlap separates the two signal classes,
\begin{equation}
\left|
\bra{\psi_0}
U_1^\dagger G_i^\dagger U_2
\ket{\psi_0}
\right|^2
\approx
\begin{cases}
1, & y_i=1, \\
0, & y_i=0.
\end{cases}
\end{equation}
For the present problem, a minimal ansatz is sufficient, and we take
\begin{equation}
U_1=I,
\qquad
U_2=S(\theta)
\end{equation}
such that the trainable squeezing on the reference copy adapts the relational measurement to the signals being discriminated.

Variational protocols based on the Loschmidt echo have previously been demonstrated using a reversing procedure to estimate the overlap between an initial and a perturbed state~\cite{liu2025variational}. Such a strategy would require reversing the relevant evolution, including the unknown nuisance interaction in the present setting. The two-copy protocol instead removes any shared nuisance unitary directly through the invariance of the overlap.

\begin{figure}[b]
\centering
\includegraphics[width=\linewidth]{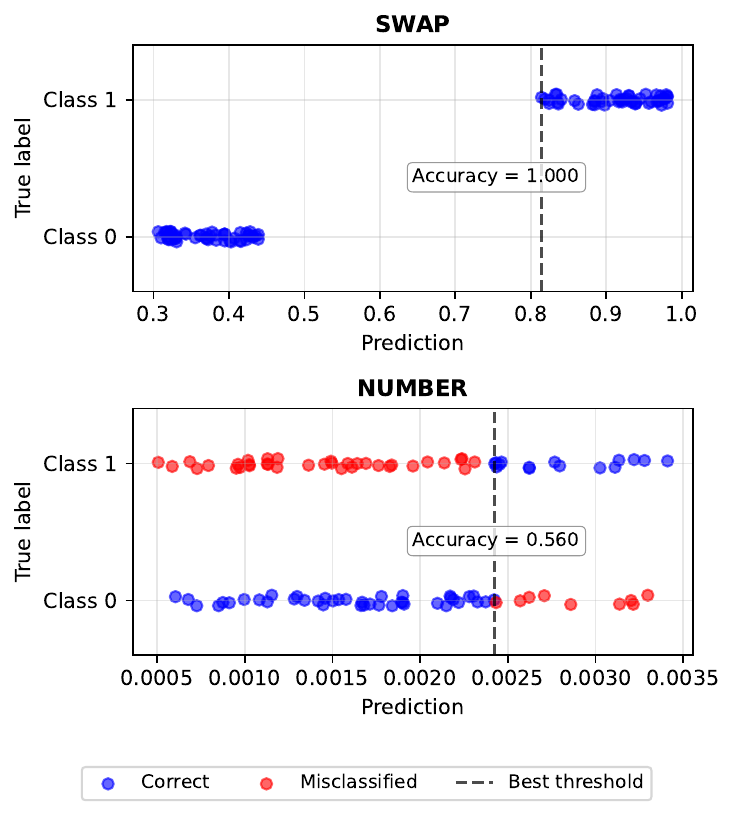}
\caption{Predictions on the test data for the sensing task using the \textsf{SWAP}-based relational model (top) and the local-observable model (bottom). Correctly classified points are shown in blue and misclassified points in red. The dashed black line denotes the classification threshold.}
\label{fig:loschmidt_scatter}
\end{figure}

Fig.~\ref{fig:loschmidt_scatter} shows the results obtained using $20$ training points over $50$ epochs. The \textsf{SWAP}-based relational model achieves perfect classification, $\mathcal{A}=1.00$, whereas the local-observable model reaches only $\mathcal{A}=0.56$. Under the assumed nuisance process, the class information is encoded in the relation between the signal and reference copies rather than in a stable phase-space orientation of either copy individually.

In practice, the signal family is likely to be at least partially known, allowing this prior information to guide the choice of a compact, physics-informed ansatz. The role of learning is then not to discover the sensing interaction from scratch, but to adapt the probe transformation to the signal strengths and discrimination task at hand. In the present example, this amounts to learning an appropriate squeezing parameter for distinguishing the two signal classes in the presence of an unknown shared bias.


\section{\label{sec:shadows}Comparing adaptive and non-adaptive quantum protocols}

The examples in the previous section rely on adaptive access to quantum data: the trainable transformations are updated during learning, and the corresponding collective observable is repeatedly evaluated on the quantum device. A natural question is whether this adaptive access is necessary, or whether the quantum data can instead be measured once and replaced by a classical representation from which the relevant quantities are estimated during subsequent training.

Classical shadows (CS)~\cite{Huang_2020cs} provide a natural framework for such ``measure-first'' approaches to quantum machine learning. In this paradigm, copies of a quantum state are measured during an initial data-acquisition stage to construct a classical shadow: a compact classical representation from which many properties of the state can subsequently be estimated. After this stage, expectation values, data features, variational losses, and other quantities required for training can be evaluated entirely classically, without further queries to the quantum device. Such protocols fall within the complexity class QESIM (Quantum-Enhanced classical SIMulation)~\cite{Cerezo_2025}, in contrast to QSIM (Quantum SIMulation), where the quantum device is queried adaptively throughout the computation, as in the relational learning architecture considered here.

Recent results have shown that shadow-based surrogates can reproduce the behavior of broad classes of trainable variational models~\cite{Bermejo_2026, Jerbi_2026}. In these settings, structural restrictions introduced to maintain trainability and avoid barren plateaus can also constrain the relevant loss landscape to a polynomial-dimensional subspace that is efficiently accessible from shadow data. This motivates the broader question of where the boundary between $\mathrm{QESIM}$ and $\mathrm{QSIM}$ lies, and in particular which trainable quantum-learning problems remain costly to reproduce using non-adaptive, measure-first protocols~\cite{kyriienko2025advantage}. In this section, we investigate this question for the relational observables considered in this work, comparing the sampling cost of shadow-based estimation with direct adaptive evaluation of the corresponding \textsf{SWAP} and \textsf{CYCLE} tests.

\subsection{Continuous-variable classical shadows}

Continuous-variable classical shadows (CVCS) based on homodyne measurements have been introduced in Refs.~\cite{gandhari2024precision, Becker_2024}. Here, we use this framework to construct a non-adaptive surrogate for the relational models considered above and to quantify the measurement cost of estimating the corresponding observables.

Following Ref.~\cite{gandhari2024precision}, we construct a CVCS estimator from $T$ homodyne measurements as
\begin{equation}
\label{eq:shadow}
\sigma_T^{(N)} = \frac{1}{T}
\sum_{t=1}^{T}
P_N F(x_{\theta_t},\theta_t)P_N,
\end{equation}
where $P_N$ projects onto a truncated Fock space and $F(x_\theta,\theta)$ is the reconstruction operator appearing in the homodyne representation,
\begin{equation}
\rho = \int d\theta \int dx_\theta \frac{1}{\pi} \bra{x_\theta}\rho\ket{x_\theta}
F(x_\theta,\theta).
\end{equation}
For each shot, a quadrature is measured at angle $\theta$, yielding an outcome $x_\theta$. The matrix elements of the corresponding reconstruction operator are
\begin{equation}
\bra{m}F(x_\theta,\theta)\ket{n}
=
e^{i(m-n)\theta}f_{mn}(x_\theta),
\end{equation}
where $f_{mn}$ are pattern functions for which efficient analytical expressions are available~\cite{LEONHARDT1996144}.

The resulting shadow can be used to estimate expectation values of selected observables. We use these estimates to construct a classical \textit{shadow dataset} $\mathcal{D}_{\mathrm{CS}} = \{(s_i,y_i)\}_{i=1}^{D}$. For each quantum data point $\ket{d_i}$, we denote by $\sigma_{T,i}^{(N)}$ the CVCS estimator constructed from $T$ homodyne measurements according to Eq.~\ref{eq:shadow}. The corresponding feature vector is defined as
\begin{equation}
\label{eq:shadow_data}
s_i = (s_i^1,\dots,s_i^M),
\qquad
s_i^m = \operatorname{Tr}\left[
\hat{O}_m \sigma_{T,i}^{(N)}
\right],
\end{equation}
where each component contains a shadow-based estimate of one of the chosen observables $\{\hat{O}_m\}_{m=1}^{M}$ for the quantum data point $\ket{d_i}$. Once these features have been extracted, all subsequent model training and inference are performed classically. The shadow construction therefore acts as a non-adaptive quantum pre-processing step that replaces repeated access to the original quantum data.

Single-mode shadow estimators can be combined to estimate multimode observables. As in other local-shadow constructions, however, the required number of measurements grows rapidly with the bodyness $k$ of the target observable~\cite{Huang_2020cs}. For the CVCS protocol considered here, an upper bound on the number of shots required to estimate a $k$-body observable $\hat{O}$ to precision $\epsilon$ with probability at least $1-\delta$ is
\begin{equation}
\label{eq:shot_upper_bound_shadows}
T = \frac{2\|\hat{O}\|_{\infty}N^{2k}\nu^{2k}}{\epsilon^2}
\left(k\log(2N) + \log\frac{1}{\delta} \right),
\end{equation}
where $\|\hat{O}\|_{\infty}$ denotes the infinity norm and $\nu^2$ is the largest single-mode variance within the support of $\hat{O}$.

For the observables of interest here, $\hat{O}$ is a permutation operator, namely \textsf{SWAP} or \textsf{CYCLE}, whose infinity norm is $1$. Hence,
\begin{equation}
T = \mathcal{O}\left(
\frac{N^{2k}\nu^{2k}}{\epsilon^2}
\log\frac{1}{\delta}
\right),
\end{equation}
up to the additional logarithmic dependence on $N$ and $k$ appearing explicitly in Eq.~\ref{eq:shot_upper_bound_shadows}. Thus, the cost of estimating the relational observable from local CV shadows grows rapidly with the number of states involved.

By contrast, direct estimation using the destructive \textsf{CYCLE} test has sampling complexity
\begin{equation}
T_{\textsf{CYCLE}} = \mathcal{O}\left(
\frac{\ln(\delta^{-1})}{\epsilon^2}
\right),
\end{equation}
with no corresponding dependence on $k$. This comparison concerns the cost of a single observable estimate for one data point. The total measurement budget of a learning protocol additionally depends on the number of data points, model evaluations, and training iterations, which we describe next.


\subsection{Shot-cost comparison}

We now compare the adaptive relational model with a non-adaptive shadow-based protocol on the hidden relationship detection task introduced in Sec.~\ref{sec:hidden_relationship}. This two-state problem provides a deliberately favorable setting for the shadow-based approach: it is the smallest non-trivial relational task considered in this work and requires estimation of only a two-body \textsf{SWAP}-based observable. As discussed above, the measurement cost of local shadow estimators becomes increasingly unfavorable as the bodyness of the target observable grows.
\begin{figure}[h!]
\centering
\includegraphics[width=\linewidth]{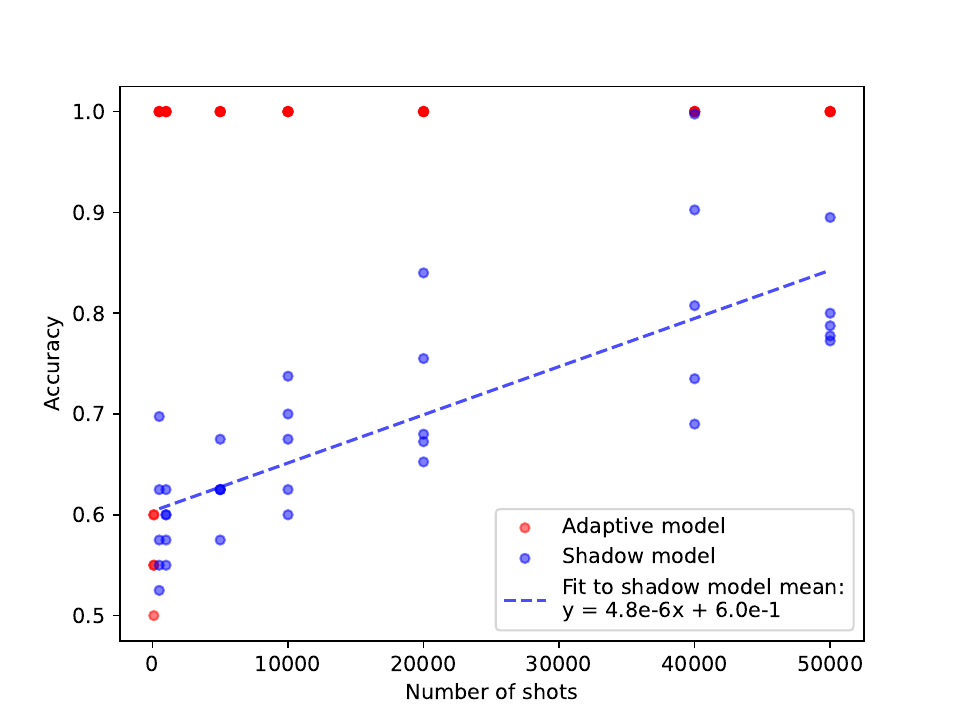}
\caption{Test accuracy of the adaptive \textsf{SWAP}-based model (red) and the non-adaptive shadow-based model (blue) on the hidden relationship dataset under finite-shot sampling. The horizontal axis shows the number of shots used for each quantum expectation-value estimate in the adaptive model and for constructing each shadow in the non-adaptive model. Five independent realizations are shown at each number of shots. The dashed line for the shadow model is a linear fit to the mean accuracy across the five realizations.}
\label{fig:shot_scaling}
\end{figure}

Fig.~\ref{fig:shot_scaling} shows the test accuracy of the adaptive and shadow-based models on $100$ unseen data points under finite-shot sampling. Both models are trained on $20$ data points over $20$ epochs. The shadow-based model is a classical linear classifier acting on the feature vectors defined in Eq.~\ref{eq:shadow_data},
\begin{equation}
\label{eq:linear_shadow_model}
f_{\mathrm{CS}}(s_i) = b+\sum_j w_j s_i^j.
\end{equation}

For the hidden relationship task, the optimal decision rule is associated with the transformed \textsf{SWAP} observable introduced in Eq.~\ref{eq:aswap}. We therefore construct shadow features corresponding to observables of the form $\hat{O}(r) = \left(R^\dagger(r)\otimes I\right) \textsf{SWAP} \left(R(r)\otimes I\right)$, for $r\in\{0.4,0.6,1,1.4,1.8\}$, with the optimal value $r^*=1$. This choice deliberately provides the shadow-based model with strong prior knowledge: the correct observable family is supplied in advance, and the optimal observable itself is included among the five candidate features. When the observables are estimated accurately, the classifier can therefore identify a sparse solution in which only the feature corresponding to $r=1$ carries non-zero weight.


Before analyzing the results, we first point out the main differences between the two approaches, which complicate a direct comparison of their total shot budgets. During both training and inference, the shadow-based protocol requires an initial collection of measurement statistics to construct the features in Eq.~\ref{eq:shadow_data}. In our simulations, the same number of shots $T$ is used for each shadow construction during training and inference per data point. This is the value reported for the shadow-based classifier in Fig.~\ref{fig:shot_scaling}. At each value on the horizontal axis, the five blue points correspond to independent realizations of the shadow measurement statistics ${(x_\theta,\theta)}$ obtained from the simulated quantum device using the number of shots reported and subsequent training of the classical model from Eq.~\ref{eq:linear_shadow_model}.

The adaptive protocol incurs additional shot costs owing to repeated use of the quantum device within the training loop. Each estimation of an expectation value for global observables requires $T_{\textsf{CYCLE}}$ shots. The total shot budget therefore depends on the length of training. In particular, it scales with the number of epochs and training points required for successful training. Moreover, gradient estimation based on the generalized shift rule~\cite{Wierichs_2022,Kyriienko2021generalized} introduces a linear shot scaling with the number of trainable parameters. In our experiments, the same number of shots is used for each \textsf{SWAP}-test evaluation across training and inference per data point. This is the value reported on the horizontal axis for the adaptive model in Fig.~\ref{fig:shot_scaling}. The red dots show the statistics of five different training runs using a given number of shots.

Based on the above, we focus on the inference shot budget, assuming that the trained model is deployed on many previously unseen data points so that the per-data-point inference cost, which affects both learning models in the same manner, becomes the relevant quantity. We observe that $T_{\textsf{CYCLE}}=500$ is sufficient for the trained adaptive model to achieve $\mathcal{A}=1.00$ on the test dataset. In contrast, the performance of the shadow-based model varies substantially across independent shadow realizations and requires a much larger shot budget. While one realization at $T=40000$ reaches near-perfect accuracy, we characterize the typical performance more robustly by fitting the mean accuracy over the five realizations at each shot budget, yielding a necessarily crude extrapolation of approximately $T\approx70000$ shots for near-perfect mean classification. At the largest shot budget investigated, the adaptive model achieves an average accuracy advantage of approximately $\Delta\mathcal{A}=0.15$ while using roughly two orders of magnitude fewer inference shots. Finally, Eq.~\ref{eq:shot_upper_bound_shadows} suggests that the measurement-cost disparity between the two approaches becomes increasingly important as the bodyness of the target relational observable grows.


\section{\label{sec:implementation}Photonic implementation}

The proposed relational-learning architecture maps naturally onto continuous-variable photonic hardware. Each of the $n$ input states undergoes a shallow trainable transformation independently, while collective processing is required only at the final \textsf{SWAP} or \textsf{CYCLE} measurement used to evaluate the corresponding $n$-state Bargmann invariant. This separation between local processing and the final collective measurement is particularly well suited to programmable time-bin architectures. Here, we outline one possible implementation on the recently reported universal photonic quantum computing architecture \emph{Clavina} \cite{yu_extensible_2026}.

\begin{figure}[t!]
\centering
\includegraphics[width=\linewidth]{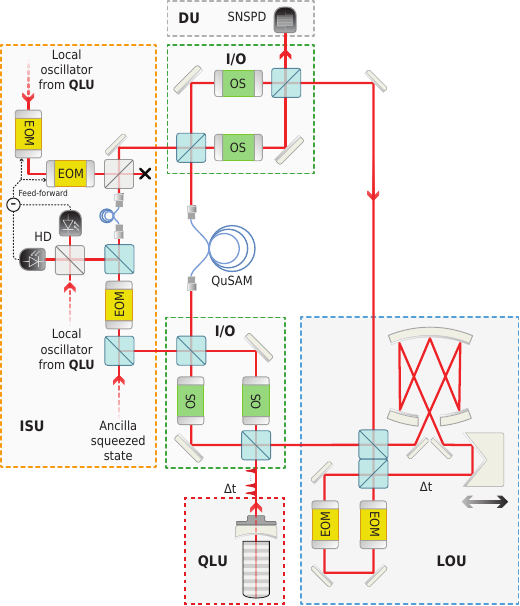}
\caption{Photonic implementation of the adaptive relational-learning architecture on \textit{Clavina}. The quantum light unit (QLU) consists of an intensity-modulated 1550~nm continuous-wave laser, which is frequency-doubled (not shown) to pump a cavity optical parametric oscillator (OPO) containing a periodically-poled potassium titanyl phosphate (PPKTP) crystal, producing time-bin SMSV states. I/O units consisting of fast, low-loss optical switches provide programmable routing between the QLU, measurement-based inline squeezing unit (ISU), linear operation unit (LOU), and detection unit (DU). High-speed electro-optic modulators (EOMs) configure the squeezing parameter in the ISU and the linear optical transformations in the LOU. The local transformations $R$, $D$, and $S$ can be applied to each time bin by repeated routing through the appropriate units. For the final collective \textsf{SWAP} or \textsf{CYCLE} measurement, the relevant time bins are routed through the LOU, which implements the required beamsplitter or Fourier interferometer, and subsequently to the DU.}
\label{fig:exp_setup}
\end{figure}

\textit{Clavina} comprises a central control unit (CU), input-output (I/O) units, a quantum light unit (QLU), a linear operation unit (LOU), an inline squeezing unit (ISU), and a detection unit (DU). The platform operates on discrete time-bin modes circulating around a loop architecture, as illustrated in Fig.~\ref{fig:exp_setup}. In this setting, the input quantum states are encoded in separate time bins, the trainable local transformations are implemented by programmable linear-optical and squeezing operations, and the final collective measurement is realized by routing the relevant time bins through the LOU before detection. High-speed optical switches (OSs) in the I/O units provide dynamic routing between the active and passive optical elements, while electro-optic modulators (EOMs) apply programmable phase shifts on a per-time-bin basis. During operation, the CU synchronizes the OSs and EOMs with long fiber delay lines acting as a quantum sequential access memory (QuSAM), allowing the same optical components to be reused across successive time-bin modes.

The hidden relationship detection and chirality-based classification tasks considered in Sec.~III require the preparation of single-mode squeezed vacuum (SMSV) input states. These states can be generated by the QLU using a low-loss cavity optical parametric oscillator (OPO), driven by an intensity-modulated, frequency-doubled continuous-wave (CW) 1550~nm laser. The initial squeezing parameter of each time bin is controlled through the pump power. With this arrangement, quadrature squeezing of approximately 10~dB has been demonstrated. The same CW laser can also supply coherent states to the circuit.

Following state preparation, each time bin is routed independently through the programmable units implementing the trainable local transformations of the model. The rotation $R(\theta)$ is mapped directly onto the LOU, where pairs of EOMs dynamically control the relative phase of each time-bin mode. When inline squeezing is not required, the I/O units route the state around the ISU, minimizing unnecessary optical loss. More general local transformations involving squeezing and displacement are implemented using the ISU.

The ISU realizes deterministic measurement-based inline squeezing using an ancillary SMSV state prepared off-line by a dedicated optical parametric amplifier. The input and ancilla fields are combined using a polarizing beamsplitter (PBS), and an EOM controls the effective squeezing level. A subsequent homodyne measurement, performed with a phase-tunable local oscillator derived from the QLU pump laser, provides the signal required for real-time feed-forward. This signal drives EOMs controlling the amplitude and phase of an additional local oscillator used to implement the displacement $D(\alpha)$. Together, these elements provide the programmable $S$ and $D$ transformations required by the local trainable circuits.

After the local transformations have been applied independently to the input states, the corresponding time bins are brought together only for the final collective measurement. The required Fourier interferometer can be realized as a linear optical circuit composed of beamsplitters and phase shifters, which \textit{Clavina}'s LOU can implement through a decomposition equivalent to a path-encoded Clements mesh. For the two-state \textsf{SWAP} test, one additional pass through the LOU implements the required 50:50 beamsplitter operation before detection. For an $n$-state \textsf{CYCLE} test, the LOU is programmed to implement the corresponding $n$-mode Fourier interferometer through a sequence of pairwise interactions between time-bin modes. While Fig.~\ref{fig:exp_setup} shows a single LOU processing core, the architecture can be extended to multiple cores. With $c$ LOU cores, $\left\lceil n/c \right\rceil$ round trips are required to implement the corresponding interaction sequence. The resulting output time bins are then routed directly to the DU, consisting of superconducting nanowire single-photon detectors (SNSPDs).


\section{\label{sec:discussion}Discussion}

Our results highlight a distinction between adaptive relational learning and measure-first approaches to quantum data, while also demonstrating the natural suitability of the former for photonic processors. The proposed model applies shallow local transformations to multiple input states before directly measuring a collective observable. By contrast, a measure-first strategy based on local classical shadows must reconstruct the same information from independently measured states. As the order of the Bargmann invariant increases, the corresponding observable becomes increasingly nonlocal, leading to an unfavorable dependence of the shadow-estimation cost on observable $k$-bodyness. The direct \textsf{CYCLE} measurement does not inherit this dependence. Our numerical comparison in the two-state setting already demonstrates a substantial difference in sampling requirements, while the scaling analysis indicates that this separation becomes more pronounced for higher-order relational observables.

Higher-order Bargmann invariants may also provide useful learning features beyond the tasks considered here. One simple example is a collective similarity measure for a set of quantum states. For pure states, $\rho_i = |\psi_i\rangle\langle\psi_i|$, the magnitude squared of the Bargmann invariant satisfies $|\Delta_n|^2 =
|\mathrm{Tr}[\rho_1\rho_2\dots\rho_n]|^2 = \mathrm{Tr}[\rho_1\rho_2]\dots \mathrm{Tr}[\rho_{n-1}\rho_n] \mathrm{Tr}[\rho_n\rho_1]$. If the fidelities between consecutive states are all $\mathrm{Tr}[\rho_i\rho_{i+1}]=1-\epsilon$, then $|\Delta_n|^2=(1-\epsilon)^n$. Thus, for fixed nonzero $\epsilon$, the cyclic invariant decreases exponentially with the number of states and provides a single global statistic of their collective similarity. Such a quantity does not reconstruct the full $O(n^2)$ pairwise similarity matrix, but may replace it in tasks where only a global assessment of the ensemble is required. More generally, the examples in this work illustrate that cyclic invariants can encode relational information beyond pairwise fidelity. Related quantities include weak values, Kirkwood-Dirac distributions, and out-of-time-ordered correlators (OTOCs)~\cite{Wagner_2024}. In many-body systems, the imaginary part of the third-order Bargmann invariant is directly related to scalar spin chirality, suggesting a connection to learning and simulation tasks in quantum magnetism~\cite{reascos2023quantum}.

We stress that the proposed architecture is not restricted to continuous-variable systems. However, photonics provides a particularly natural setting because independently prepared quantum states can undergo programmable local transformations before being combined in a final interferometric measurement. This structure allows relatively simple trainable circuits to access genuinely multi-state observables. Photonic \textsf{SWAP} measurements for quantum-data machine-learning pipelines have already been demonstrated outside the setting considered here~\cite{li2026machinelearningquantumdata}, and Sec.~\ref{sec:implementation} shows how the full architecture considered in this work can be mapped onto the recently reported \textit{Clavina} platform~\cite{yu_extensible_2026}. An important next step is therefore to identify experimentally relevant tasks in which adaptive access to quantum data provides a measurable advantage over measure-first strategies.


\section{\label{sec:conclusions}Conclusions}

We developed an adaptive relational learning framework for multi-state quantum data based on Bargmann invariants. We applied independent trainable transformations to each input state and evaluated their relations through collective \textsf{SWAP} and \textsf{CYCLE} measurements. Using continuous-variable (CV) photonic examples, we demonstrated hidden relationship detection, geometric-phase classification, and sensing, achieving perfect accuracy on quantum-data learning tasks that are not directly accessible to classical machine learning models.

We also compared the adaptive model with a measure-first approach based on continuous-variable classical shadows. We showed that estimating higher-order relational observables from local shadows incurs an unfavorable dependence on observable bodyness, whereas direct collective measurements avoid this scaling. Already in the simplest two-state setting, our numerical results demonstrated an orders-of-magnitude reduction in the required shot budget for the adaptive approach. Finally, we mapped the model onto a programmable time-bin photonic architecture, providing a concrete route toward experimental implementation.


\section*{Acknowledgements}
M.\,J. and O.\,K. are grateful to Leonardo Novo, Stefano Scali, and Chukwudubem Umeano for productive discussions on the subject. O.\,K. M.\,J., and R.\,B.\,P. acknowledge the support from UK EPSRC award under the Agreement No. EP/Z53318X/1 (QCi3 Hub). S.\,Y. acknowledges support from the United Kingdom Research and Innovation (UKRI) Guarantee Postdoctoral Fellowship (project: EP/Y029631/1). R.\,B.\,P. acknowledges support from the UKRI Future Leaders Fellowship (project: MR/W011794/1) and The Royal Society (project:  RG\textbackslash R2\textbackslash 232514).


\appendix

\section{Details of local-observable models used}
\label{app:local_models}

As mentioned in Sec.~\ref{sec:use_cases}, Figs.~\ref{fig:rot_data_results},~\ref{fig:chiral_dataset_predictions},~\ref{fig:loschmidt_scatter} compare the performance of the relational models in each of the presented use cases with adaptive models based on the local observable $f(\ket{d_i}) = \bra{d_i}U^\dagger(\Theta)\hat{n}_1U(\Theta)\ket{d_i}$, where $\hat{n}_1$ is the number operator on the first mode. We use the conventions
\begin{align}
R(\theta) &= \exp\left(i\theta, a^\dagger a\right), \\
S(r,\varphi) &= \exp\left[\tfrac{1}{2}\left(r e^{-i\varphi} a^2 - r e^{i\varphi} a^{\dagger 2}\right)\right], \\
\mathrm{BS}(\theta,\varphi) &= \exp\left[\theta\left(e^{i\varphi} a_1 a_2^\dagger - e^{-i\varphi} a_1^\dagger a_2\right)\right],
\end{align}
where $\varphi = 0$ unless stated otherwise and $S(r) \equiv S(r,0)$. The Fock-space truncation was verified for each run by checking that the population of the highest Fock level remained below $10^{-3}$. Below, we detail the exact ansatz used for each problem.

For the hidden relationship task in Sec.~\ref{sec:hidden_relationship}, the ansatz consists of a single rotation on the second mode, with $\Theta = (\theta_1)$:
\begin{equation}
U(\Theta) = I \otimes R(\theta_1).
\end{equation}

For the chirality task in Sec.~\ref{sec:chirality}, the ansatz consists of a single-mode squeezer on the third mode followed by a tritter, with $\Theta = (\theta_1, \theta_2)$:
\begin{equation}
U(\Theta) = T(\theta_2)\left(I \otimes I \otimes S(\theta_1)\right),
\end{equation}
where the tritter is the parametrized three-mode interferometer
\begin{equation}
T(\theta_2) = \mathrm{BS}_{12}(\theta_2, -\tfrac{2\pi}{3}) \mathrm{BS}_{23}(\tfrac{\pi}{4}, \tfrac{2\pi}{3}) \mathrm{BS}_{12}(\theta_2, 0),
\end{equation}
with $\mathrm{BS}_{ij}$ acting on modes $i$ and $j$, and the same trainable angle $\theta_2$ shared by both outer beamsplitters.

Finally, for the sensing task in Sec.~\ref{sec:sensing_task}, the trainable gates act both before and after the data interaction $G_i$. The full state entering the measurement is
\begin{equation}
\ket{d_i} = U(G_i,\Theta, \phi_i)\ket{0,0},
\end{equation}
with $U(G_i,\Theta, \phi_i) = \left(S(\theta_3) \otimes I\right)\mathrm{BS}(\theta_2, 0) (R(\phi_i)\otimes R(\phi_i)) (I\otimes G_i) \left(S(\theta_1) \otimes I\right)$ and $\Theta = (\theta_1, \theta_2, \theta_3)$.

\bibliography{main}
\end{document}